# Hyperdense Coding Modulo 6 with Filter-Machines

Vince Grolmusz *

**Abstract**

We show how one can encode $n$ bits with $n^{o(1)}$ "wave-bits" using still hypothetical filter-machines (here $o(1)$ denotes a positive quantity which goes to 0 as $n$ goes to infity). Our present result - in a completely different computational model - significantly improves on the quantum superdense-coding breakthrough of Bennet and Wiesner (1992) which encoded $n$ bits by $\lceil n/2 \rceil$ quantum-bits. We also show that our earlier algorithm (Tech. Rep. TR03-001, ECCC, ftp://ftp.eccc.uni-trier.de/pub/eccc/reports/2003/TR03-001/index.html) which used $n^{o(1)}$ muliplication for computing a representation of the dot-product of two $n$-bit sequences modulo 6, and, similarly, an algorithm for computing a representation of the multiplication of two $n \times n$ matrices with $n^{2+o(1)}$ multiplications can be turned to algorithms computing the exact dot-product or the exact matrix-product with the same number of multiplications with filter-machines. With classical computation, computing the dot-product needs $\Omega(n)$ multiplications and the best known algorithm for matrix multiplication (D. Coppersmith and S. Winograd, Matrix multiplication via arithmetic progressions, J. Symbolic Comput., 9(3):251–280, 1990) uses $n^{2.376}$ multiplications.

## 1 Introduction

It is one of the first tasks in any undergraduate information theory or computer science course to show that general $n$-bit sequences cannot be compressed to a shorter sequence or cannot be encoded by less than $n$ bits. The proof of these results are based on the fact that any injective image of a $2^n$-element set must contain exactly $2^n$ elements.

However, using some fascinating physical phenomena and different models of computation, superdense coding is possible. Bennet and Wiesner [1], using Einstein-Podolski-Rosen entangled pairs, showed that $n$ classic bits can be encoded by $\lceil n/2 \rceil$ quantum bits. Note, that this result is optimal in the quantum model.

Here we describe an algorithm for encoding $n$ bits with $n^{o(1)}$ "wave-bits", using a different model, the filter-machines, to be defined in the next section.

## 2 Preliminaries

### 2.1 The dot-product

We have defined the *alternative*, and the *0-a-strong* and the *1-a-strong* representations of polynomials in [3] and [4]. Since we need only the notation of 1-a-strong representation here, we reproduce here only that definition. Note also, that for prime or prime-power moduli, polynomials and their representations (defined below), coincide.

*Department of Computer Science, Eötvös University, Budapest, Pázmány P. stny. 1/C, H-1117 Budapest, Hungary; E-mail: grolmusz@cs.elte.hu

**Definition 1** ([4]) *Let $m$ be a composite number $m = p_1^{e_1} p_2^{e_2} \cdots p_\ell^{e_\ell}$. Let $Z_m$ denote the ring of modulo $m$ integers. Let $f$ be a polynomial of $n$ variables over $Z_m$:*

$$f(x_1, x_2, \ldots, x_n) = \sum_{I \in \{0,1,2,\ldots,d\}^n} a_I x_I,$$

*where $a_I \in Z_m$, $x_I = \prod_{i=1}^n x_i^{\nu_i}$, where $I = \{\nu_1, \nu_2, \ldots, \nu_n\} \in \{0, 1, 2, \ldots, d\}^n$. Then we say that*

$$g(x_1, x_2, \ldots, x_n) = \sum_{I \in \{0,1,2,\ldots,d\}^n} b_I x_I,$$

*is a 1-a-strong representation of $f$ modulo $m$, if $\forall I \in \{0, 1, 2, \ldots, d\}^n$ $\exists j \in \{1, 2, \ldots, \ell\}$: $a_I \equiv b_I \pmod{p_j^{e_j}}$, and, furthermore, if for some $i$, $a_I \not\equiv b_I \pmod{p_i^{e_i}}$, then $a_I \equiv 0 \pmod{m}$.*

**Example 2** *Let $m = 6$, and let $f(x_1, x_2, x_3) = x_1 x_2 + x_2 x_3 + x_1 x_3$, then $g(x_1, x_2, , x_3) = x_1 x_2 + x_2 x_3 + x_1 x_3 + 3x_1^2 + 4x_2$ is a 1-a-strong representation of $f$ modulo 6.*

In other words, for modulus 6, in the 1-a-strong representation, the non-zero coefficients of $f$ are correct for both moduli in $g$, but the zero coefficients of $f$ can be non-zero either modulo 2 or modulo 3 in $g$, but not both.

In [4] we proved the following theorem:

**Theorem 3** *Let $m = p_1 p_2$, where $p_1 \neq p_2$ are primes. Then a degree-2 1-a-strong representation of the dot-product $f(x_1, x_2, \ldots, x_n, y_1, y_2, \ldots, y_n) = \sum_{i=1}^n x_i y_i$ can be computed as the homogeneous bilinear form:*

$$\sum_{j=1}^t \left( \sum_{i=1}^n b_{ij} x_i \right) \left( \sum_{i=1}^n c_{ij} y_i \right) \tag{1}$$

*where $b_{ij}, c_{ij} \in \{0, 1\}$ and $t = \exp(O(\sqrt{\log n \log \log n})) = n^{o(1)}$.*

In other words, we have shown, that instead of the usual dot-product $\sum_{i=1}^n x_i y_i$ we can compute a polynomial of the form

$$\sum_{i=1}^n x_i y_i + 3g(x, y) + 4h(x, y) \tag{2}$$

where both $g$ and $h$ has the following form: $\sum_{i \neq j} a_{ij} x_i y_j$, $a_{ij} \bmod 6 \in \{0, 1\}$, and no term $x_i y_j$ appears in both $f$ and $g$.

## 3 The Filter-Machine

In this short communication we give the definition only for modulo 6; for other non-prime-power composites the definition can easily be generalized.

**Definition 4** *Let $G(z) = f(z) + 3g(z) + 4h(z)$ be a polynomial of $m$ variables $z = (z_1, z_2, \ldots, z_m)$, where the coefficient of every monomial in $f$ is 1 modulo 6, and no monomial appears in two of the polynomials $f, g, h$ with non-zero coefficients modulo 6. Then $M$ is a mod 6 filter-machine for polynomial $G(z)$, if for inputs $G(z)$ and $\zeta \in \{0, 1, 2, 3, 4, 5\}^m$ $M$ returns in one step the value*

$$f(\zeta) \bmod 6.$$

## 3.1 Notes on realization and motivation

Let us consider polynomial $G(z)$, and suppose that we can increment the value of polynomials $f$, $g$, and $h$ independently from each other. Then the period of $3g$ is 2, the period of $4h$ is 3, while the period of $f$ is 6, all seen modulo 6. So if we were able to filter out the shorter period (that is, the higher frequency) "waves" then we were able to compute $f(\zeta)$. Note, that for doing this we may need to sustitute values from $Z_6$ instead just bits into the polynomials.

Note, that machine $M$ does not need access to the actual values of the variables of the polynomials $g$ and $h$, it just needs access to their periodically changed values.

After filtering $g$ and $h$ out, it asks for the value (somehow similarly as the quantum machines perform an observation) of $G(\zeta)$, reduced by this "filtering", which is just $f(\zeta)$.

Let us see the most important example: the dot-product. Let $G$ be the polynomial of (2). Suppose that we would like to retrieve the value of $x_1 = \xi_1 \in \{0, 1\}$. Now, if we plug in $y_1 = 1, y_2 = y_3 = \cdots = y_n = 0$, then we shall get $x_1 + 3(x_{i_1} + x_{i_2} + \cdots + x_{i_s}) + 4(x_{j_1} + x_{j_2} + \cdots + x_{j_k})$, where $i_u \ne j_v, u = 1, 2, \ldots, s, v = 1, 2, \ldots, k$. Now, $M$ assumed to have access to some values of $x_{i_1} + x_{i_2} + \cdots + x_{i_s}$ in order to filter them out, since their periodicity is at most 2 modulo 6; and also to some values of $x_{j_1} + x_{j_2} + \cdots + x_{j_k}$ to filter them out, since their periodicity is at most 3 modulo 6. Note again, that their $xi_i$ values are not needed at this phase, and, also, that typically, it is not enough to substitute 0 and 1 in the variables, $0, 1, 2, 3, 4, 5$ may be needed.

After identifying the higher frequency terms, $M$ filter them out, and returns the value of $f$, which is $\xi_1$ in our case. Note, that we ask only here for the value of a variable.

## 4 Hyperdense coding

Polynomial (2) can be computed in form (1). Let us consider an $x$, and let us compute $X = (X_1, X_2, \ldots, X_t)$, where

$$X_j = \sum_{i=1}^{n} b_{ij} x_i \bmod 6 \qquad (3)$$

that is, simply a homogeneous linear function of $x$, determined by (1), for $j = 1, 2, \ldots, t$. However, for a given substitution $x = \xi \in \{0, 1\}^n$ it is not enough to store the mod 6 numbers (3) (since different $\xi$'s will lead to the same $X_j$ values, because $t = n^{o(1)} < n$), but rather, we need to store $X_j$'s in a form which facilitates the independent periodicity (or frequency) testing of the filter-machine.

### 4.1 Hiperdense encoding algorithm

The encoding is done by linear transformations (3).

### 4.2 Hiperdense decoding algorithm

Suppose that we would like to decode $\xi_1$. Then plug in $y = (1, 0, 0, ...0)$ into (1). Then, from (2), we get

$$\xi_1 + 3g + 4h,$$

and the $3g + 4h$ sum can be cancelled out by the filter-machine (for example, as it was hinted in subsection 3.1).

## 5 Dot-product, matrix-vector product, matrix-product

We gave algorithms in [4] with $n^{o(1)}$, $n^{1+o(1)}$, and $n^{2+o(1)}$ multiplication for computing the 1-a-strong representation of the dot-product, matrix-vector product, matrix-product, respectively. Using filter-machines, these representations can be turned to the computing of the exact values with 1, $n$, $n^2$ further filter-machine operations, respectively.

The best known algorithm today for matrix-multiplication was given by Coppersmith and Winograd [2], requiring only $n^{2.376}$ multiplications.